\documentclass[12pt]{article}
\usepackage{hyperref}
\usepackage{amsmath,amssymb,accents}
\usepackage{latexsym}
\newcommand{\Mat}[1]{{{\boldsymbol{#1}}}}

\def\Log{\mathrm{Log}\,}
\def\be{\begin{equation}}
\def\ee{\end{equation}}
\def\bea{\begin{eqnarray}}
\def\eea{\end{eqnarray}}
\def\bc{\begin{center}}
\def\ec{\end{center}}
\def\bi{\begin{itemize}}
\def\ei{\end{itemize}}
\def\bs{\begin{frame}}

\def\dd{\operatorname{d}}

\def\noi{\noindent}
\begin{document}

\title{Continuum dynamics and the electromagnetic field in the scalar ether theory of gravitation}

\author{
Mayeul Arminjon$^{1,2}$\\
$^1$ \small\it  \ Univ. Grenoble Alpes, Lab. 3SR, F-38000 Grenoble, France. \\
$^2$ \small\it CNRS, Lab. 3SR, F-38000 Grenoble, France.\\
\small E-mail: Mayeul.Arminjon@3sr-grenoble.fr
} 

\date{}
			     
\maketitle

\begin{abstract}
An alternative, scalar theory of gravitation has been proposed, based on a mechanism/interpretation of gravity as being a pressure force: Archimedes' thrust. In it, the gravitational field affects the physical standards of space and time, but motion is governed by an extension of the relativistic form of Newton's second law. This implies Einstein's geodesic motion for free particles only in a constant gravitational field. In this work, equations governing the dynamics of a continuous medium subjected to gravitational and non-gravitational forces are derived. Then, the case where the non-gravitational force is the Lorentz force is investigated. The gravitational modification of Maxwell's equations is obtained under the requirement that a charged continuous medium, subjected to the Lorentz force, obeys the equation derived for continuum dynamics under external forces. These Maxwell equations are shown to be consistent with the dynamics of a ``free" photon, and thus with the geometrical optics of this theory. However, these equations do not imply local charge conservation, except for a constant gravitational field.\\

\noi {\it Keywords:} Alternative theories of gravitation; preferred reference frame; curved spacetime; Maxwell equations; charge conservation\\

\noi Published in {\it Open Physics} {\bf 14}, 2016, 395--409. This work is licensed under the Creative Commons Attribution-NonCommercial-NoDerivs 3.0 License.

\end{abstract}
\section{Introduction}
\subsection{General motivation}

Since Einstein, most physicists admit that physics definitely obeys the 
principle of relativity. Poincar\'e, although he formulated the principle 
in its full generality \cite{Poincare1904}, and although he explored many of its 
consequences in great detail \cite{Poincare1905,Poincare1906}, considered its validity as a possibility which has to be tested by experiments. Lorentz and Poincar\'{e} always 
reserved the opposite possibility: that some physical phenomenon might 
contradict the principle of relativity. (Kaufmann's experiment indeed seemed 
to exhibit such phenomenon.) This was the physical justification for their 
attitude regarding the ether --- an attitude which had also philosophical 
reasons. Following Builder \cite{Builder1958a,Builder1958b} and J\'{a}nossy \cite{Janossy1965}, the ``Lorentz-Poincar\'{e}" ether interpretation of special relativity (SR) has been thoroughly discussed and has been proved to be entirely consistent; see e.g. Prokhovnik \cite{Prokhovnik1993} and references therein. Since Lorentz invariance is true in the Lorentz-Poincar\'{e} ether interpretation, it is empirically indistinguishable from standard SR \cite{Prokhovnik1967}, except for the following fact. In contrast with the case in standard SR, the limit velocity $c$ has not an ``absolute" status in this interpretation. This is because, in it, the Poincar\'e-Einstein conventional simultaneity is regarded as ``true" only in one reference frame (the ``ether"). That is, a preferred simultaneity exists in this interpretation: the Poincar\'e-Einstein simultaneity in the ether frame. Therefore, according to this interpretation, a signal velocity $v>c$ would not violate causality  (\cite{O3}, Note 6). Apart from this fact, that ether cannot be detected. So the role played by the ether in that interpretation may be qualified as ``metaphysical".\\

But SR does not describe gravitation. An alternative theory of gravitation has been proposed, that offers a mechanism/interpretation allowing to understand gravity as being a pressure force: Archimedes' thrust \cite{O3,A18}. 
\footnote{\
A new version (hereafter v2) of that theory has been built in Refs. \cite{O3,A35}. Unless mentioned otherwise, what is discussed in this paper applies to v1 and v2. 
}
It is thus a scalar theory but an original one (see below), and it is a preferred-frame theory. I.e., it violates the local Lorentz invariance because its ether has physical effects, but it is ``relativistic" in the sense of Will \cite{Will1993}: ``in the limit as gravity is `turned off', the nongravitational laws of physics reduce to the laws of special relativity". Thus, the Lorentz-symmetry violations are in the gravitational sector, and are therefore very small (see below). Note that extensions of general relativity (GR) that break Lorentz invariance have been proposed, e.g. \cite{JacobsonMattingly2001}. Note also that, according to this theory, as in the Lorentz-Poincar\'e version of SR and for just the same reason given above, a signal velocity $v>c$ would not violate causality. Thus that theory could survive, in contrast to general relativity, if signal velocities $v>c$ were observed --- but $v>c$ in the ether frame is forbidden for an usual mass particle, i.e. one having $m_0 $ real (and positive); see Eq. (\ref{m(v)}). As exposed in Sect. 2.1 of Ref. \cite{A28}, other motivations for this theory come: (A) from the wish to concile quantum physics with the theory of gravitation, and (B) from some difficulties in GR itself (despite its impressing successes): (i) The unavoidable singularities. (ii) The necessity to regard diffeomorphic Lorentzian spacetimes $(\mathrm{V},\Mat{\gamma})$ and $(\mathrm{V}',\Mat{\gamma}')$ as equivalent, which is handled by adding a gauge condition (four scalar ``coordinate conditions") to the Einstein equations \cite{FriedrichRendall2000}. (iii) The need for dark matter and dark energy. Regarding (A): it has been found recently that the curved-spacetime Dirac energy operator has a non-uniqueness problem, and that a most satisfying solution of it can be implemented if and only if the spacetime metric has the form postulated in v2 of the present theory, Eqs. (\ref{gamma_0i=0}) and (\ref{Space metric-v2}) \cite{A48}. Note that this form distinguishes a preferred reference frame. Regarding (B): for Point (i) the present theory predicts a ``bounce" instead of a singularity for both the gravitational collapse of a dust sphere \cite{A18} and also around the past high-density state of the universe implied by the cosmological expansion \cite{A28}. (ii) In the present scalar theory, the spacetime manifold is given and there is no need for a gauge condition. (iii) For the dark matter problem, we have a plausibility argument: the preferred-frame effects should have a greater effect at large scales, because the large orbital times allow these effects to accumulate. As to the dark energy problem: the theory necessarily predicts an acceleration of the cosmic expansion \cite{A28}. 
\subsection{Brief summary of the theory}

(See Sect. 2 of Ref. \cite{B26} for an extended summary, and see Sect. \ref{Dynamics_Particle} below for a self-consistent exposition of what is needed here.) This is a scalar theory written in a preferred reference frame. The scalar field $\beta $ has two roles: (i) It determines the relation between the ``physical" spacetime metric (that which is more directly related with measurements by physical clocks and rods) and a ``background" Minkowski metric, of which the spacetime manifold $\mathrm{V}$ is assumed to be endowed \cite{O3}. (ii) It generates a gravity acceleration, Eq. (\ref{Def vector g}) below \cite{A16}. Indeed, {\it dynamics is governed by a generalization of Newton's second law to curved spacetime.} That new dynamics implies geodesic motion for free test particles only in a static gravitational field \cite{A16}. The equation for the scalar field is extremely simple: for v2, it is 
\be\label{wave eqn v2}
\square \psi = (4\pi G/c^2) \sigma,
\ee
with $\psi\equiv-\Log \beta $, $\square$  the flat-spacetime wave operator, and $\sigma\equiv T^{00}$, the energy component of the energy-momentum tensor \cite{A35}. This equation is valid in coordinates adapted to the preferred reference frame, and with $x^0=cT$ where $X\mapsto T$ is a preferred time map on the spacetime manifold $\mathrm{V}$. The equation for the scalar field will not be used in the present work, however. As to the assumption relating the physical and background metric, it will be used merely in Subsect. \ref{Newton continuum} and in Sect. \ref{Wave vs ray} --- for which the same conclusions would be drawn with the different assumption that was set in v1. Therefore, {\it in this paper, it is essentially only the original dynamics of the theory that is relevant.}

\subsection{Current state of the experimental check}

Note first that already its original dynamics (detailed in Sects. \ref{Dynamics_Particle} and \ref{Dynamics_Continuum}) implies that this theory is different from all known scalar theories. 
\footnote{\
This includes that of Ref. \cite{WattMisner1999}, even though, just like in the latter scalar theory, the metric assumed in the v2 version of the present theory is the same as Ni's metric \cite{Ni1972}  --- as noted in Ref. \cite{B26}.
}
Much work has been done to test the v1 version of this theory. (See a detailed summary as Sect. 4 in Ref. \cite{O2}.) The experimental tests of GR, the currently accepted theory of gravitation, are mostly in a weak gravitational field. The so-called ``parameterized PN formalism" (see Will \cite{Will1993} and references therein) does not apply to the present theory for the following reason: geodesic motion being valid only in the static case, this is not a ``metric theory" \cite{A19}. Since the tests of GR involve very accurate experiments \cite{Will1993}, there is a need for a very clean post-Newtonian (PN) approximation scheme. We have developed an asymptotic scheme of post-Newtonian approximation, in accordance with the general principles of asymptotic analysis \cite{A23}. Using this scheme, one shows first that v1 has the correct Newtonian limit \cite{A23} (and v2 also \cite{A35}). At the second approximation (1PN), one has to take into account the motion of the mass center of the gravitating system, assumed isolated --- the solar system, say --- with respect to the ether frame. That motion may, for many purposes, be envisaged as a uniform translation, the velocity vector ${\bf V}$ of which is not \textit{a priori} known. Its magnitude may yet be estimated to be at most of the order 300 km/s from various astronomical arguments, if the ether frame is assumed to coincide with the average rest frame of matter. It has been proved \cite{A19} that the uniform velocity $V$ has no effect on the motion of photons, at the 1PN approximation which is used to confront GR with the experimental observations of gravitational effects on light rays \cite{Will1993}. In fact, the 1PN predictions of this theory for photons are completely indistinguishable from the standard 1PN predictions of GR, even though in this theory photons do not exactly follow the ``null" geodesics of the spacetime metric \cite{A19}. The v1 version of the theory passed a number of other tests, e.g. regarding celestial mechanics in the solar system \cite{B21} and binary pulsar energy loss by emission of gravitational waves \cite{A34} (in both cases, considering a system of extended bodies and accounting for a velocity $V \ne 0$ of its mass center). Note also that this theory {\it predicts} an acceleration of the cosmic expansion \cite{A28}. But v1 has been discarded by a significant violation of the weak equivalence principle (WEP), which has been found to occur for extended bodies at the point-particle limit. That violation occurs due to the fact that the spatial metric is anisotropic, in much the same way as is the standard form of the Schwarzschild metric \cite{B26}.\\

That violation of the WEP does not occur any more for v2 with its isotropic spatial metric, Eq. (\ref{Space metric-v2}) below \cite{A35}. In particular, in the static spherical case, the 1PN approximation of the spacetime metric is the same as the standard 1PN metric of GR \{\cite{A35}, Eq. (88)\}. The celestial-mechanical tests should be redone with v2 but, as discussed in Ref. \cite{O2}, \S 4.6, there is a lot of specialized parameter adjustment in celestial mechanics, whose test is hence less decisive than is sometimes believed. In view of the static spherical case and the improvement w.r.t. v1 regarding the WEP, one may expect that v2 should improve over v1 in celestial mechanics. Also, recall that for v2 the field equation is the exact flat-spacetime wave equation (\ref{wave eqn v2}). Since for v1 the latter wave equation was got at the relevant post-Minkowskian approximation \cite{A34}, one expects that similar results will be obtained regarding the binary pulsar energy loss as with v1. For light rays, it has been checked in Ref. \cite{A35} that also for v2 the 1PN predictions of this theory for photons are indistinguishable from the standard 1PN predictions of GR. This means that the gravitational effects on electromagnetic rays: the gravitational redshift, the deflection of light, and the time delay, are predicted by this theory as they are in GR. The geodetic precession measured by the Gravity Probe B (GP-B) experiment \cite{Everitt2011} is calculated from geodesic motion in the Schwarzschild metric \cite{Schiff1960} (correcting for the Earth's oblateness \cite{Breakwell1988}), so that the same prediction is got from the present theory. As to the frame-dragging (Lense-Thirring) effect, it is not predicted by this theory, but its confirmation by GP-B is far less precise than for the geodetic precession. Moreover, the asymptotic scheme of PN approximation predicts effects of the self-rotation of gravitationally active bodies, also for GR \cite{A36}.\\

In view of the number and the complexity of the experimental tests of gravitation, it is clearly unfeasible for the proponent of a truly alternative theory (still more than for the proponent of a mere extension of GR) to check all of them. Hence, despite many efforts and many good points, this theory has currently the status of a tentative theory.

\subsection{Motivation and aim of the present paper}

The different dynamics as compared with GR implies that the extension of the laws of 
non-gravitational physics from SR to the situation with gravitation cannot 
be done as straightforwardly in this theory as in GR. In GR and in other 
``metric theories" of gravitation, the extension of such a law is done simply 
by formally substituting the curved metric to the flat Minkowski metric of 
SR into the covariant expression of this law (e.g. Stephani \cite{Stephani1982}, Will \cite{Will1993}). This formal substitution means actually a modification of the law by its 
coupling with gravitation \cite{Stephani1982}, the mathematical expression of this coupling 
being thus obtained in an automatic way. The central equation of SR that one 
adapts to curved spacetime is that for continuum dynamics, thus 
obtaining the well-known equation $T^{\mu \nu }_{\, ;\nu } =0$ for the energy-momentum tensor {\bf T}. 
\footnote{\ 
Greek indices vary from 0 to 3, Latin ones from 1 to 3 (spatial 
indices). Semi-colon means covariant derivative associated with the 
physical~spacetime metric, the latter being denoted by $\Mat{\gamma} $. 
Indices are raised and lowered with the help of this metric, unless 
explicitly mentioned otherwise.
}
As a consequence of this equation, a dust (a continuum made of free test 
particles) has a geodesic motion, and this explains why the above procedure 
cannot be used in the present theory. In this theory, one proceeds in the 
reverse way: the equation for dust is deduced from Newton's second law and 
rewritten in terms of the energy-momentum tensor. The obtained 
expression is assumed to be valid for any kind of continuum or system of fields, characterized by the expression of tensor {\bf T} as a function of the state variables 
\cite{A15,A20}. This assumption is justified by the mass-energy equivalence and the 
universality of the gravitational force. An interesting consequence of the 
obtained dynamical equation is that, for a perfect fluid, the mass 
conservation is obtained as a limiting behaviour at low pressures or in a 
weak and slowly varying gravitational field. Whereas, under high pressure in a strongly varying field, the theory predicts that matter is produced or destroyed \cite{A20}.\\

{\it In this paper,} we extend the previous method (induction from a dust to a 
general behaviour) to the situation where non-gravitational forces are 
present. We apply this to the case where the non-gravitational force is the 
Lorentz force, and obtain the gravitationally-modified Maxwell equations of 
the present theory. This leads us to a really discriminating prediction of the theory as compared with GR. We also examine the link between those modified Maxwell 
equations and photon dynamics as governed by Newton's second law for a 
light-like test particle --- that is, we examine the transition from wave 
optics to geometrical optics in the presence of gravitation. We find that, 
as in GR, this transition is provided by the case of a ``null"  electromagnetic field. In GR, the null fields come into play {\it via} the discontinuities equations for an electromagnetic shock wave (e.g. Lichnerowicz \cite{Lichne1962}, Synge \cite{Synge1964}, de Felice {\&} Clarke \cite{deFelice-Clarke1990}). In the present theory, null fields occur simply because a null field behaves like a dust of photons, hence our extension of Newton's second law applies.

\section{Dynamics of a test particle under gravitational and 
non-gravitational forces}\label{Dynamics_Particle}

Let us first list the assumptions used in the present work. As in GR, it is assumed: \\

\noi {\it a}) that our space and time measurements may be arranged so as to be described by a metric $\Mat{\gamma} $ with ($+ - - -)$ signature on a 4-dimensional manifold $\mathrm{V}$ (the spacetime). 
\footnote{\
See Ref. \cite{A16}, \textsection 2.1, for a discussion of this assumption that emphasizes its compatibility with a preferred-frame theory. However, point ({\bf ii}) there, i.e. the formal transcription rule from SR to GR, is not relevant to this theory, as it has been pointed out in the introduction of the present paper. The transcription of the expression of the non-gravitational force will be examined in Section \ref{Lorentz & Maxwell} in the case of the Lorentz force.
}\\

\noi {\it b}) That a continuous medium or a physical field is defined by the expression of its energy-momentum tensor {\bf T}. The latter has to be {\it a symmetric spacetime tensor field whose value ${\bf T}(X)$ at an event $X$ depends only on the values at $X$ of the smooth functions (field components) that characterize the continuous medium/physical field which is being considered} (\cite{Fock1964}, \S $31^\ast$). The tensor {\bf T} is obtained in the following way. As shown by Fock: in SR, the former condition in italics, plus the demand that the divergence of {\bf T} vanish, determine uniquely the expression of {\bf T} --- at least in the concrete examples examined by him and by coworkers. These include the {\bf T} tensor for the electromagnetic field (\cite{Fock1964}, App. B) and for the perfect fluid (\cite{Fock1964}, App. C). Still in SR, this expression is then rewritten in generally-covariant form (\cite{Fock1964}, \S 46). Finally, the expression of {\bf T} in a general coordinate system in a general spacetime is got immediately, by just substituting the curved spacetime metric $\Mat{\gamma }$ for the Minkowski metric in the generally-covariant expression of {\bf T} that is valid in SR. See e.g. \cite{Fock1964}, \S 60, for the perfect fluid, and \cite{L&L}, Eq. (94.8), for the electromagnetic field. Note that this procedure {\it does not use a Lagrangian} from which the equations of motion of the matter fields be derived through the action principle.
\footnote{\
In the preferred-frame theory considered in this paper, one may postulate a matter Lagrangian and hence a matter action that are invariant only under the coordinate changes having the form (\ref{spatial change +f(t)}). The definition of the Hilbert tensor ${\bf T}_\mathrm{H}$ as the variational derivative of the Lagrangian density \{e.g. \cite{L&L}, Eq. (94.4)\} is still applicable. But this definition is got from studying the variation of the action under a small change of the coordinate system; see a precise statement as Theorem 1 in Ref. \cite{A53}. Due to Eq. (\ref{spatial change +f(t)}) here and to the boundary condition to be verified by the allowed small changes of the coordinate system, one may show that they are all zero. This makes that definition irrelevant. It implies also that the dynamics of ${\bf T}_\mathrm{H}$ is not constrained by this restricted invariance of the action (whereas the general invariance of the action, when it takes place, determines this dynamics to be $T^{\mu\nu}_{\ \,;\nu}=0$, e.g. \cite{L&L}). In addition, the proof of tensoriality of ${\bf T}_\mathrm{H}$ (Theorem 2 in Ref. \cite{A53}) can be adapted to this case, but the tensorial transformation is then got merely for the changes (\ref{spatial change +f(t)}). Therefore, we shall {\it not} postulate a Lagrangian and the action principle. 
}
 \\

In addition to Assumptions ({\it a}) and ({\it b}), merely three features of that theory will be used in the present work --- except for Eq. (\ref{Space metric-v2}) which is used in Subsect. \ref{Newton continuum} and in Sect. \ref{Wave vs ray}. The first two features are common with GR, even though they are not widely used in GR. The third feature is specific to that theory.\\

\noi {\bf i}) 
In the spacetime manifold $\mathrm{V}$, a {\it reference fluid} $\mathcal{F}$ is basically a three-dimensional network of reference points. Each reference point is defined by its world line, which has to be time-like for a physically admissible reference fluid \cite{Cattaneo1958}. Thus, $\mathcal{F}$ may be defined by the associated unit tangent 4-vector field $\Mat{U}=\Mat{U}_\mathcal{F}$  \cite{Cattaneo1958, RodriguesCapelas2007}: the reference world lines are the integral curves of $\Mat{U}_\mathcal{F}$. A coordinate system is said to be {\it adapted} to some reference fluid $\mathcal{F}$, iff each reference world line has constant space coordinates $x^i\ (i=1,2,3)$ \cite{Cattaneo1958}.  In adapted coordinates, a reference point, or ``point bound to $\mathcal{F}$", may hence be specified by the vector ${\bf x}\equiv (x^i) \in \mathbb{R}^3$. In Ref. \cite{A52}, it is proved that adapted coordinates do exist for any ``normal" non-vanishing vector field on $\mathrm{V}$, and a rigorous definition of the {\it space manifold} associated with a reference fluid (defined by a normal vector field $\Mat{U}$) is formulated. The space manifold is the set $\mathrm{N}_\Mat{U}$ of the reference world lines (that set being endowed with a metrizable topology and an atlas of compatible charts). Thus, the ``physical space" can be regarded as being the set $\mathrm{N}_\Mat{U}$ of the reference world lines; it depends on the reference fluid which is considered. In any such reference fluid, we have a spatial metric $\Mat{g}= \Mat{g}_\mathcal{F}$ (it too depends on $\mathcal{F}$)  \cite{L&L, Cattaneo1958}. That metric depends in general of the time coordinate, i.e., the reference fluid is deformable, hence the name ``reference fluid". Moreover, at any point bound to $\mathcal{F}$, we have a local time  $t_{\mathbf{x}}$  \cite{L&L, Cattaneo1958}. This is the proper time measured at some fixed point ${\bf x}$ bound to $\mathcal{F}$. \\

\noi {\bf ii}) 
We assume that there is a {\it preferred} reference fluid $\mathcal{E}$, with four-velocity vector field $\Mat{U}_\mathcal{E}$, which is globally synchronized \cite{A16,L&L}, i.e., there is a global spacetime coordinate system ($x^\mu )$ which is adapted to $\mathcal{E}$, and in which the components of the spacetime metric verify
\footnote{\
This condition alone does not specify a unique reference fluid, even less a unique coordinate system \cite{A16,L&L}. The preferred character of $\mathcal{E}$ appears with the dynamical equation (\ref{Newton 2nd law}) with (\ref{Def vector g}), which is covariant only under the coordinate transformations (\ref{spatial change +f(t)}).
}
\be\label{gamma_0i=0}
\gamma_{0 i}=0.
\ee
It implies in particular that the local time $t_{\bf x}$, related to the coordinate time $t\equiv x^0/c$ of such coordinates by
\be\label{Local time}
 \frac{\dd t_{\bf x}}{\dd t} = \beta \equiv \sqrt {\gamma_{00} } ,
\ee
is synchronized along any trajectory \cite{L&L}. \\

\noi {\bf iii}) 
We assume that, in this reference fluid $\mathcal{E}$, the equation of motion for a test particle is the following one (``Newton's second law"):
\be\label{Newton 2nd law}
{\bf F} + (E/c^{2}) {\bf g} =  D{\bf P}/Dt_{\mathbf{x}},
\ee
with ${\bf F}$ the non-gravitational force and ${\bf g}$ the gravity 
acceleration, and where $E$ is the ``purely material energy" of the test 
particle (i.e., not accounting for the potential energy in the gravitational 
field). For a mass point, this is defined as $E = m(v) c^{\, 2}$, with
\be\label{m(v)}
m(v) \equiv  m_0\gamma_{v} ,\qquad \gamma_{v}\equiv 1/\sqrt{1-(v^{ 
2}/c^{2})}. 
\ee
Here $v$ is the modulus of the 3-velocity of the test particle relative to the reference fluid $\mathcal{E}$.  That velocity ${\bf v}$ is measured with the local time $t_{\mathbf{x}}$ and its modulus $v$ is defined with the space metric $\Mat{g}$ (\textit{we mean $\Mat{g}= 
\Mat{g}_{\mathcal{E}\, } $ and $\Mat{U}=\Mat{U}_\mathcal{E}$ from now on}) :
\be\label{Def v}
v^{\, i} \equiv \frac{\dd x^i}{\dd t_{\mathbf{x}}}  \equiv \frac{1}{\beta} \frac{\dd x^i}{\dd t}, \quad v \equiv  [\Mat{g}({\bf v}, {\bf v})]^{1/2} =(g_{ij} v^i\, v^j)^{1/2}.
\ee
For a photon, we define $E = h\nu $, $h$ being Planck's constant and $\nu $ the 
frequency as measured with the local time of the momentarily coincident 
observer of the reference fluid $\mathcal{E}$: $\nu\equiv \dd n/\dd t_{\bf x}$ with $n$ the number of periods. Further, in Eq. (\ref{Newton 2nd law}), ${\bf P }$ is the momentum, given by 
\be
{\bf P} \equiv  (E/c^{\, 2}) {\bf v}. 
\ee
Finally, in Eq. (\ref{Newton 2nd law}), $D/Dt_{\mathbf{x}}$ is the relevant derivative of a time-dependent vector in the space manifold $\mathrm{N}_\Mat{U}$ endowed with the time-dependent metric $\Mat{g}$ (and rescaled 
to the local time $t_{\mathbf{x\, \, }}$ as for $v^{\, i}$ in Eq. (\ref{Def v}), i.e., $D{\bf P}/Dt_{\bf x}\equiv  (1/\beta )D{\bf P}/Dt$). Compelling arguments \cite{A16} give the unique expression
\be\label{Def D/Dt}
D{\bf P}/Dt \equiv D_{0\, }{\bf P}/\textit{Dt} + (1/2) {\bf t.P}, 
\quad {\bf t} \equiv  \Mat{g}^{-1}{\rm {\bf .}}\frac{\partial {\kern 1pt}{\kern 
1pt}{\kern 1pt}\Mat{g}}{\partial \,t}.
\ee
In particular, this ensures that Leibniz' rule is satisfied for the 
derivation of the scalar product: $\dd (\Mat{g}({\bf u}_1,{\bf u}_2))/\dd t=\Mat{g}({\bf u}_1,D{\bf u}_2/Dt)+\Mat{g}(D{\bf u}_1/Dt,{\bf u}_2)$. In Eq. (\ref{Def D/Dt})$_{1}$, $D_{0\, 
}{\bf P}/Dt$ is the absolute derivative relative to the ``frozen" space 
metric $\Mat{g}$ of the time $t$ ($t\equiv x^{\, 0}/c)$ where the derivative is to be 
calculated:
\be\label{D_0 P/Dt}
\left ( \frac{D_0{\bf P}}{Dt} \right )^i \equiv \frac{\dd P^i}{\dd t} + \Gamma ^i_{jk} P^j\, \frac{\dd x^k}{\dd t},
\ee
with $\Gamma ^i_{jk}$ the Christoffel symbols associated with the metric $\Mat{g}$ of the time $t$. From the definition of the proper time $\tau $ along a general trajectory:
\be
\dd s^2 =c^2 \dd \tau ^2 = \gamma _{\mu \nu }\, \dd x^\mu \, \dd x^\nu,
\ee
one gets the following relation between $\tau $, the local time $t_{\bf x}$, and the coordinate time $t\equiv x^0/c$:
\be\label{tau-t_x-t}
\frac{\dd \tau }{\dd t} = \frac{\dd \tau }{\dd t_{\bf x}} \frac{\dd t_{\bf x}}{\dd t} = \frac{\beta }{\gamma _v}.
\ee

\vspace{2mm}
It turns out that ``Newton's second law" (\ref{Newton 2nd law}) is compatible with the 
formulation of motion in GR, provided a peculiar (velocity-dependent) form 
of the gravity acceleration  ${\bf g}$ is assumed \cite{A16}. Hence, that 
part of our assumption ({\bf iii}) which is specific to the present theory is in fact only the assumed form for ${\bf g}$: In that theory, ${\bf g}$ is a space vector given by
\be\label{Def vector g}
{\rm {\bf g}}=-\,\frac{c^{2}}{2}\frac{\mbox{grad}_{g} {\kern 1pt}\gamma 
_{{\kern 1pt}00} }{\gamma_{{\kern 1pt}00} }
=-\,c^{2}\,\frac{\mbox{grad}_{g} {\kern 1pt}\beta }{\beta },
\quad
\left( {\mbox{grad}_{g} \beta } \right)^{i}\equiv g^{ij}\beta_{,j} \;,
\left( {g^{ij}} \right)\equiv \left( {g_{ij}} \right)^{-1}.
\ee
This expression and Newton's second law (\ref{Newton 2nd law}) are covariant under coordinate changes that both leave the reference fluid unchanged and keep true the  
synchronization condition $\gamma_{0\,i} =0$, i.e.
\be\label{spatial change +f(t)}
x'^0 = \varphi (x^0), \quad x'^i= \psi^i(x^1,x^2,x^3).
\ee
However, for Eq. (\ref{Newton 2nd law}), this covariance is true only if one assumes that the non-gravitational force ${\bf F}$ is an invariant spatial vector (field) under any change (\ref{spatial change +f(t)}), i.e., the components $F^i$ are contravariant under the change $x'^i= \psi^i(x^1,x^2,x^3)$, 
\footnote{\
Then ${\bf F}$ can be rigorously defined as a vector field on the space manifold $\mathrm{N}_\Mat{U}$ --- more exactly, in the case that it depends on the coordinate time $t=x^0/c$, as a one-parameter family $({\bf F}_t)$ of vector fields on $\mathrm{N}_\Mat{U}$.
}
and in addition they are invariant under the allowed changes of the time coordinate, $x'^0 = \varphi (x^0)$. 
This condition is imposed upon ${\bf F}$, because ${\bf g}, {\bf v}, {\bf P}$ and $D {\bf P}/Dt_{\mathbf{x}}$ do possess it. (The local time interval $\dd t_{\mathbf{x}}$ is an invariant for these transformations, and so is the energy $E$.)\\

The expression (\ref{Def vector g})$_{1}$ for {\bf g} comes up naturally from a (semi-heuristic) interpretation of gravity as being Archimedes' thrust due to the macroscopic pressure in a perfectly fluid ``ether" \cite{A18,O3}. But the same expression (\ref{Def vector g})$_{1}$ can be {\it derived} by demanding that (i) the metric field $\Mat{\gamma} $ should be a spatial potential for a space vector {\bf g}, and (ii) the law of motion (\ref{Newton 2nd law}) should imply geodesic motion for ``free" test particles in a \textit{static} metric \cite{A16}. For the case of a free test particle, i.e. ${\bf F} = {\bf 0}$ in Eq. (\ref{Newton 2nd law}), this law implies the following energy balance \cite{A15}:
\begin{equation}
\label{eq3}
\frac{\dd \left( {E{\kern 1pt}\beta } \right)}{\dd t}=E\frac{\partial {\kern 
1pt}\beta }{\partial {\kern 1pt}t}.
\end{equation}
This is true, again, for both mass points and photons. As a consequence of 
this equation, it is obvious that the total energy of the material test 
particle, including its ``potential" energy in the gravitational field, must 
be defined as $e_{\, m} \equiv E\beta $, which is a constant for a 
constant gravitational field \cite{A15}. The proof of Eq. (\ref{eq3}) for a mass point 
(pp. 42--43 in Ref. \cite{A15}) is a modification of the elementary method used in 
classical mechanics to derive the (potential plus kinetic) energy equation 
in a force field deriving from a variable potential --- with some complications 
due to relativistic mechanics with a variable metric. Now, in the case 
where a non-gravitational force {\bf F} is present as in Eq. (\ref{Newton 2nd law}), it is straightforward to modify the proof and to get
\be\label{dE/dt} 
\frac{\dd \left( {E{\kern 1pt}\beta } \right)}{\dd t}=E\frac{\partial {\kern 
1pt}\beta }{\partial {\kern 1pt}t} + \beta^{\, 2} {\bf F}{\bf .v}, \qquad
{\bf F}{\bf .v} \equiv  \Mat{g}({\bf F}, 
{\bf v}) \equiv  g_{ij}  F^i v^j.
\ee
Using the  synchronization condition, $\gamma_{0\,i}  =0$, the 
expression of the 4-acceleration of a free mass point has been deduced from 
the equation of motion (\ref{Newton 2nd law}) and the energy balance (\ref{eq3}) in Ref. \cite{A16}: this is the result of some algebra with Christoffel symbols. It is again 
straightforward to adapt this calculation to the case ${\bf F}\ne {\bf 0}$, thus deducing from Eqs. (\ref{Newton 2nd law}) and (\ref{dE/dt}) the following 
expression:
\be\label{4-accel}
A^0 = \frac{1}{2 \beta ^2}\,g_{jk,0}\,U^j\,U^k +  \frac{\gamma _v}{\beta }\frac{{\bf F} {\bf .v}}{m_0\,c^3}, \quad A^i = \frac{1}{2}\,g^{ij} g_{jk,0}\,U^0\,U^k +\gamma _v\frac{ F^i}{m_0\,c^2}.
\ee
Here ${\bf U}$, with components $U^\mu $, is the 4-velocity of the mass point (not to be confused with $\Mat{U}$, the 4-velocity field of the preferred reference fluid that defines the space manifold $\mathrm{N}_\Mat{U}$). Thus:
\be 
U^\mu \equiv \frac{\dd x^\mu}{\dd s}, \quad A^\mu \equiv \left(\frac{\Delta  {\bf U}}{\Delta s} \right)^\mu,
\ee
$\Delta /\Delta s$ being the absolute derivative relative to the spacetime metric $\Mat{\gamma} $. Like Eqs. (\ref{Newton 2nd law}) and (\ref{Def vector g}), Eqs. (\ref{dE/dt}) 
and (\ref{4-accel}) are covariant under the transformations (\ref{spatial change +f(t)}), assuming again that the external force {\bf F} is a invariant spatial vector field in the sense defined after Eq. (\ref{spatial change +f(t)}).

\section{Dynamics of a continuum under gravitational and non-gravitational forces}\label{Dynamics_Continuum}

\subsection{Induction from a dust}\label{From dust}

We seek the dynamical equation satisfied by the energy-momentum tensor {\bf T} in the present theory. (See Assumption ({\it b}) at the beginning of Sect. \ref{Dynamics_Particle}.) That equation will determine how a given kind of continuum, whose behaviour is well-identified in SR, couples to gravitation in the present theory. Here, we consider the case where a non-gravitational external force field is present, in addition to the gravitation. If this non-gravitational force field is considered \textit{given,} the same dynamical equation must apply to any kind of continuum: this is the way to express the mass-energy equivalence and the universality of the gravitation force in the framework of Assumption ({\it b}) of Sect. \ref{Dynamics_Particle}. (However, the non-gravitational force depends 
actually on the considered continuum, of course: only the gravitational 
force is universal.) Therefore, the dynamical equation may be derived for a 
dust. Dust is a continuum made of coherently moving, non-interacting 
particles, each of which conserves its rest mass --- so that the dynamical 
equation for mass test particles translates immediately into that for the 
dust continuum. In doing so, we have to substitute in Eqs. (\ref{Newton 2nd law}) and (\ref{dE/dt}) the constant rest-mass of a ``substantial" volume element: $\delta m_{0} = \rho_{0} \delta V$, and the external force on this element: $\delta {\bf F}= {\bf f} \delta V$, for $m_{0}$ and ${\bf F}$ respectively, with $\rho_{0}$ the rest-mass density in the preferred reference fluid and ${\bf f}$ the density of the non-gravitational external force. Both densities are evaluated with respect to the physical volume measure: 
\be\label{delta V}
\delta V = \sqrt{ g}\, \dd x^{1}\dd x^{2}\dd x^{3},  \qquad g \equiv \mathrm{det}
(g_{ij}).
\ee
In Ref. \cite{A20}, the equation has been derived for a dust in the 
absence of (non-gravitational) external force. Here, we add an external 
force, hence we have to use the expression (\ref{4-accel}) of the 4-acceleration. The {\bf T} tensor in energy units is given, for a dust of mass points, by \cite{Fock1964}:
\be\label{T dust}
T^{\, \mu \nu } = \rho ^\ast \,c^{\, 2}U^{\, \mu }U^{\, \nu } .
\ee
(Of course the 4-velocity ${\bf U}$, with components $U^\mu$, is now a field.) Mass conservation: $(\rho^\ast U^{\nu })_{\,;\nu}=0$ \cite{Fock1964} is true for that dust, by definition, whence:
\be
T^{\mu \nu}_{\ \,;\nu} = \rho^\ast c^2\, U^{\mu }_{\,;\nu}\,U^\nu = \rho ^\ast  c^2\,A^\mu.
\ee
Multiplying Eq. (\ref{4-accel}) by $\rho ^\ast  c^2$ and accounting for this, we get for a dust of mass points:
\be\label{Eq T-f}
T^{0 \nu}_{\ \,;\nu} =b^0({\bf T})+\frac{{\bf f.v}}{c\beta},
\qquad T^{i \nu}_{\ \,;\nu} =b^i({\bf T})+ f^i,
\ee
where
\be\label{b^mu}
b^0({\bf T}) \equiv \frac{1}{2\beta^2}\,g_{ij,0}\,T^{ij},
\quad b^i({\bf T}) \equiv \frac{1}{2}\,g^{ij}g_{jk,0}\,T^{0k},
\ee
which depend linearly on {\bf T}. We assume that this dynamical equation in the presence of a field of external force density {\bf f}, Eq. (\ref{Eq T-f}) with (\ref{b^mu}), is true for a general continuum. It is covariant under the transformations (\ref{spatial change +f(t)}) (for an invariant field {\bf f}; note that the volume element (\ref{delta V}) is invariant under (\ref{spatial change +f(t)})). 

\subsection{Newton's second law for a general continuum}\label{Newton continuum}

In the foregoing subsection, passing through the intermediary of the 
4-acceleration to get the dynamical equation (\ref{Eq T-f}) for a dust, we induced the validity of that equation for a general continuum. As an alternative to this way of doing, it is interesting also to directly apply Newton's second law (\ref{Newton 2nd law}) and the energy equation (\ref{dE/dt}), extending them from a point test particle to a general continuous medium. This will be necessary for Sect. \ref{Wave vs ray}, and it also enables one to identify the internal forces in the continuum. Let us first consider a {\it dust of mass points.} In the continuous case, we treat a small ``substantial" volume $\delta V$ of the dust continuum as a test particle. The density $\rho_{0}\equiv \delta m_0/\delta V$ is related to the proper rest-mass density $\rho ^\ast $ 
by \cite{Fock1964}:
\be\label{rho_0 vs rho ast}
\rho_{0} = \gamma_{v} \rho ^\ast .
\ee
(See also Ref. \cite{A15}, Eq. (4.28).) The {\bf T} tensor is given by Eq. (\ref{T dust}) and, because $\gamma _{0\,i} =0$, we have $U_{0} = \gamma_{00\, }U^{\, 0}$ while, in view of (\ref{tau-t_x-t}), $U^0 \equiv \dd t/\dd \tau = \gamma _{v}/\beta = \gamma_{v}/\sqrt{ \gamma_{00}}$. Using (\ref{rho_0 vs rho ast}), we hence find that the amount of ``purely material energy" contained in a volume element $\delta V$ is, for such a dust:
\be\label{delta E}
\delta E \equiv  \delta m_0 \gamma_{v} c^{2} = \rho_{0} \gamma_{v}\, c^{2}\delta V = \rho ^\ast  \gamma _{v}^{2} c^{2} \delta V = \rho ^\ast  c^2 \left(\frac{\dd t_{\bf x}}{\dd \tau} \right )^2\,\delta V = \rho ^\ast  c^2 \, U^0 U_{0}\,\delta V = T^0_{\ \, 0} \delta V.
\ee
Therefore, to express the mass-energy equivalence and the universality of the gravitation force consistently within Assumption ({\it b}) of Sect. \ref{Dynamics_Particle}, we have to \textit{define} the amount of ``purely material energy" for a {\it general continuum} as $\delta E = T^0_{\ \, 0} \delta V$. Hence, the law of motion (\ref{Newton 2nd law}) must be written for a volume element of a continuous medium as:
\be\label{Newton 2nd law continuum}
{\bf f'} \delta V +\frac{T^0_{\ \, 0}}{c^2}  \delta V {\bf g} 
=\frac{1}{\beta}\,\frac{D}{Dt}\left (\frac{T^0_{\ \, 0}}{c^2\,\beta} {\bf u} \delta V \right ), \quad {\bf u}\equiv \frac{\dd {\bf x}}{\dd t} = \beta {\bf v}, \quad {\bf f'}\equiv \frac{\delta {\bf F}}{\delta V} . 
\ee
In this expression, ${\bf f '}$ is the density of the \textit{total} non-gravitational force over an element of the continuum, thus including the \textit{internal} forces such as stresses (though reexpressed as a volume density force). Also, it is understood that, as indicated above, the volume element is ``substantial". I.e., it follows the motion of the continuum; that may include deformation, so that $\delta V$ may depend on the time coordinate. To go further, it is convenient to utilize the ``bimetric" nature of the theory. We will use the assumption of v2, according to which the space metric is 
\be\label{Space metric-v2}
\Mat{g}=\beta ^{-2} \Mat{g}^0,
\ee
with $\Mat{g}^0$ an {\it Euclidean} metric on the space manifold $\mathrm{N}_\Mat{U}$ \cite{O3,A35}. This implies that the ``physical" measure of the volume element, $\delta V \equiv \sqrt{ g}\,\dd x^1\,\dd x^2\,\dd x^3$, is related to the Euclidean measure $\delta V^{0}$ of this same element by
\be\label{dV vs dV^0}
\delta V = \delta V^{0}/\beta^3,\qquad \delta V^{\, 0} \equiv \sqrt{ g^{0}}\,\dd x^1\,\dd x^2\,\dd x^3,
\ee
where $g^0\equiv $ det ($g^{0\, }_{ij})$ is associated with the metric $\Mat{g}^{0}$. \{Ref. \cite{A35}, Eq. (19). In the v1 version of the theory, the metric assumption was different and led to $\delta V = \delta V^{0}/\beta$, Eq. (4.29) in Ref. \cite{A15}. With v1, the same agreement is found between the corresponding equations as the one found below between Eqs. (\ref{Newt 2nd continuum-4}) and (\ref{Dyn T i-2}) and between Eqs. (\ref{Eq T-f})$_1$ and (\ref{Energy free dust-2}).\} The metric $\Mat{g}^{0}$ is assumed \textit{constant} in the preferred reference fluid, i.e., $g^0_{ij,0} =0$ in any coordinates ($x^{\mu })$ which are adapted to the preferred reference fluid. However, Eq. (\ref{Space metric-v2}) (and Eq. (\ref{dV vs dV^0})$_{1}$ as well) is not stable under  a change of the time coordinate. Hence, this assumption privileges a particular time coordinate $x^0 =cT$, where $T$ is called the ``absolute time" \cite{A35}. We shall note $\tilde{\beta} \equiv (\sqrt{ \gamma _{00}})_{x^0=cT}$. The constancy of $\Mat{g}^{0}$ implies that, for a volume element which follows the motion of the continuum, we have
\be\label{d dV /dt}
\dd (\delta V^{0})/\dd t = \delta V^0\, \mathrm{div}_{0} 
{\bf u},\quad \mathrm{div}_{0}  {\bf u} \equiv  \mathrm{div}_{\Mat{g}^0}\, 
{\bf u} =(u^{ i}\,\sqrt{g^0})_{\,, i}/\sqrt{g^0}.
\ee 
\{Note that Eq. (\ref{d dV /dt}) is covariant under any change (\ref{spatial change +f(t)}).\} Inserting Eqs. (\ref{dV vs dV^0}) and (\ref{d dV /dt}) into Eq. (\ref{Newton 2nd law continuum}) leads to
\be\label{Newt 2nd continuum-2}
{\bf f'} + \frac{T^0_{\ \, 0}}{c^2} {\bf g} = \tilde{\beta}^2\,\frac{D}{DT} \left (\frac{T^0_{\ \, 0}}{c^2\,\tilde{\beta}^4} \,{\bf u}\right ) +\frac{T^0_{\ \, 0}}{c^2\,\tilde{\beta}^2} {\bf u}\, \mathrm{div}_{0}  {\bf u} \qquad  (x^0 = cT ).
\ee
It shall be simpler for the sequel to rewrite this space vector equation in covector form, lowering the indices of spatial vectors with the physical space metric $\Mat{g}$. This commutes \cite{A16} with the 
\textit{D/Dt} derivative, defined for a spatial covector ${\bf w}^\ast $ by  \cite{A16}:
\be\label{D/Dt covector}
\frac{D{\bf w}^\ast}{Dt} \equiv \frac{D_0{\bf w}^\ast }{Dt} -\frac{1}{2} {\bf t.w}^\ast, \quad i.e. 
\left(\frac{D{\bf w}^\ast}{Dt}\right)_i = \frac{\partial w_i}{\partial t} + w_{i\mid j} \frac{ \dd x^j}{\dd t} -\frac{1}{2} t^k_{\ \, i}\,w_k,
\ee
where $w_{i\mid j}$ denotes the covariant derivatives (thus here those corresponding to the spatial metric $\Mat{g}$) of the covector ${\bf w}^\ast $. Moreover, the following identity applies in the case that
${\bf w}^\ast$ has the form ${\bf w}^\ast = \rho  {\bf u}^\ast$ :
\be\label{cov rho u}
\left(\rho u_i \right )_{\mid k}\,u^k = \left(\rho u_i \right )_{, k}\,u^k - \frac{1}{2}g_{jk,i}\, \rho u^j u^k,
\ee
where $u^j \equiv  g^{jk} u_{k}$. Using these two properties, one rewrites Eq. (\ref{Newt 2nd continuum-2}) as
\be\label{Newt 2nd continuum-3}
\tilde{\beta}^{-2} f'_i + \tilde{\beta}^2\rho g_i = \frac{\partial \left( \rho u_i \right )}{\partial T}+\left (\rho u_i  \right )_{,k} u^k - \frac{1}{2}\left (\frac{\partial g_{ij} }{\partial T}\,\rho u^j+g_{jk,i}\rho u^j u^k \right ) + \rho u_i \mathrm{div}_0 {\bf u},
\ee
where $f'_{i}\equiv  g_{ij} f'^j$ and the like for $g_{i}$, and where 
\be\label{Def rho v2}
\rho \equiv \frac{T^0_{\ \, 0}}{c^2\,\tilde{\beta}^4} = \frac{T^{0 0}}{c^2\,\tilde{\beta}^2} \qquad (x^0=cT).
\ee
Due to (\ref{d dV /dt})$_2$, this rewrites in spatial coordinates such that $g^{0}_{\ ,j} = 0$ (e.g. Cartesian coordinates for the Euclidean metric $\Mat{g}^0$) as:
\be\label{Newt 2nd continuum-4}
\tilde{\beta}^{-2} f'_i + \tilde{\beta}^2\rho g_i = \frac{\partial \left( \rho u_i \right )}{\partial T}+\left (\rho u_i  u^k \right )_{,k}  - \frac{1}{2}\left (\frac{\partial g_{ij} }{\partial T}\,\rho u^j+g_{jk,i}\rho u^j u^k \right ).
\ee

\vspace{2mm}
Let us compare this with Eq. (\ref{Eq T-f})$_2$. In any coordinates adapted to the reference fluid $\mathcal{E}$ and such that $x^0 =cT$, one has from (\ref{dV vs dV^0}): $\gamma \equiv $ det($\gamma 
_{\lambda \nu }) =-\gamma_{00}.g = -\tilde{\beta}^2.(g^0/\tilde{\beta}^6) = -g^0 \tilde{\beta}^{-4}$. Using the identity
 \be\label{Tmu nu;nu mixed}
T_{\mu\ \ ;\nu}^{\ \,\nu}=\frac{1}{\sqrt{-\gamma}} \left ( \sqrt{-\gamma}\,T_{\mu}^{\ \,\nu}\right )_{,\nu}-\frac{1}{2}\gamma_{\lambda \nu ,\mu } T^{\lambda \nu } ,
\ee
and noting that $\gamma _{0i}=0$ and $\gamma _{jk}=-g_{jk}$, Eq. (\ref{Eq T-f})$_2$ may hence be rewritten, in spatial coordinates such that $g^{0}_{\ ,j} = 0$, as 
\be\label{Dyn T i}
\tilde{\beta}^2 \left (\frac{T_i^{\ \,\nu} }{\tilde{\beta}^2} \right)_{,\nu }= -f_i + c^2\rho \tilde{\beta}^3 \tilde{\beta}_{,i}-\frac{1}{2}\left ( g_{ij,0}T^{0j} + g_{jk,i}T^{jk}\right ) \quad   (x^0= cT ). 
\ee
(Here $f_i\equiv g_{ij}\,f^j$.) From (\ref{Def vector g}), we get
 \be\label{g covector}
g_{i} \equiv  g_{ij} g^j = -c^2\tilde{\beta}_{, i }/\tilde{\beta}. 
\ee
Define $u^0 \equiv \dd x^0/\dd t=c$, so that $u^\mu=\dd x^\mu /\dd t$. Considering again a dust of mass points, we get from (\ref{tau-t_x-t}) and (\ref{T dust}): 
\be\label{T dust-1}
T^{\mu \nu }\equiv  \rho ^\ast c^2 U^\mu  U^\nu = \rho ^\ast \frac{\gamma _v^2}{\tilde{\beta }^2}\,u^\mu u^\nu.
\ee
But, in view of (\ref{delta E}), we have $\rho ^\ast \frac{\gamma _v^2}{\tilde{\beta }^2} c^2 =T^{0 0}$. Therefore, we get from the definition (\ref{Def rho v2}):
\be\label{T dust-2}
T^{\mu \nu }= \rho\,\tilde{\beta }^2\,u^\mu u^\nu.
\ee
It follows that 
\be\label{T dust-3}
T_i^{\ \,0} = - c \tilde{\beta }^2\,\rho u_{i}, \quad T_i^{\ \,j} = - \tilde{\beta }^2\,\rho u_{i\, }u^{\, j}.
\ee
Using (\ref{g covector}) and (\ref{T dust-3}) in Eq. (\ref{Dyn T i}) and multiplying by $\tilde{\beta}^{-2}$, it becomes:
\be\label{Dyn T i-2}
\partial _T \left (-\rho u_i \right) -\left(\rho u_i \,u^j\right)_{,j}= -\tilde{\beta}^{-2}\,f_i -\rho \tilde{\beta}^2 g_i-\frac{1}{2}\left ( (\partial _T\,g_{ij})\rho u^j + g_{jk,i}\,\rho u^j\,u^k \right ).
\ee
Thus, for a dust of mass points, Eqs. (\ref{Newt 2nd continuum-4}) and (\ref{Dyn T i}) coincide if $f_{i} = f'_i$, but obviously this is not true for a general tensor {\bf T}. The reason is simple: in the direct transposition (\ref{Newton 2nd law continuum}) of Eq. (\ref{Newton 2nd law}), leading to Eq. (\ref{Newt 2nd continuum-4}), ${\bf f '} $ is the density of the \textit{total} non-gravitational force, as we mentioned. In contrast, Eq. (\ref{Eq T-f}), and then Eq. (\ref{Dyn T i}), have been obtained by induction from a dust, thus a continuum with \textit{zero} internal force, to a general tensor {\bf T}. Hence {\bf f} in Eqs. (\ref{Eq T-f}) and (\ref{Dyn T i}) is indeed the density of the \textit{external }non-gravitational force. By making Eqs. (\ref{Newt 2nd continuum-4}) and (\ref{Dyn T i}) coincide for a given expression of the {\bf T} tensor and a given external force density {\bf f}, we may identify the volume density of the internal forces in a general continuum, i.e., ${\bf f''} \equiv  {\bf f ' }- {\bf f}$.\\

As to the energy equation, we merely write the transposition of Eq. (\ref{eq3}), 
thus to a dust in the absence of non-gravitational external force. In view 
of Eq. (\ref{delta E}), this is
\be\label{Energy free dust-0}
\frac{\dd}{\dd t}\left ( T^0_{\ \,0}\,\delta V \beta \right )=T^0_{\ \,0}\, \delta V \frac{\partial \beta}{\partial t}.
\ee
With the help of Eqs. (\ref{dV vs dV^0}) and (\ref{d dV /dt}), the l.h.s. rewrites as
\be\label{energy-rate free dust}
\frac{\dd}{\dd T}\left ( \frac{T^0_{\ \,0}}{\tilde{\beta }^2 }\,\delta V^0 \right )=\frac{\dd}{\dd T}\left ( T^{0 0}\,\delta V^0 \right )= \left (\frac{\dd T^{0 0}}{\dd T}+T^{0 0 }\,\mathrm{div}_0{\bf u} \right ) \delta V^0 \quad (x^0=cT),
\ee
thus (\ref{Energy free dust-0}) is
\be
\frac{\dd \epsilon }{\dd T}+\epsilon \,\mathrm{div}_0{\bf u} = \frac{\epsilon }{\tilde{\beta }}\,\frac{\partial  \tilde{\beta }}{\partial T},\quad \epsilon \equiv (T^{0 0})_{x^0=cT}.
\ee
This may be rewritten (in any spatial coordinates) as
\be\label{Energy free dust} 
\frac{\partial \epsilon }{\partial T} + \mathrm{div}_0 \left(\epsilon {\bf u}\right) = \frac{\epsilon }{\tilde{\beta }}\,\frac{\partial  \tilde{\beta }}{\partial T} .
\ee
For a dust, we have from (\ref{T dust-2}):

\be\label{Current free dust}
T^{0 0}\,u^i = cT^{0 i}.
\ee
With this equation, Eq. (\ref{Energy free dust}) becomes:
\be\label{Energy free dust-2} 
 \frac{\partial T^{0 0} }{\partial T} + \mathrm{div}_0 \left( c T^{0 j} \partial _j \right) =  T^{0 0} \frac{\partial \Log \tilde{\beta}}{\partial T} \quad (x^0=cT).
\ee
This is equivalent to Eq. (\ref{Eq T-f})$_1$ (\cite{A35}, Eqs. (24)--(26) and Note 5).\\

In classical physics, there is just one external force field in addition to 
the gravitation: namely, the electromagnetic (Lorentz) force.

\section{Lorentz force and Maxwell equations in a gravitational field}\label{Lorentz & Maxwell}

The electromagnetic (e.m.) field is defined by the spacetime tensor $\Mat{F}$. We 
assume that $\Mat{F}$ derives from a 4-potential {\bf A}:
\be\label{Def F}
F_{\mu \nu } \equiv  A_{\nu ,\mu } - A_{\mu, \nu } = A_{\nu;\mu } - A_{\mu ; \nu },
\ee 
which is equivalent to assuming that $\Mat{F}$ is antisymmetric ($F_{\mu \nu 
} = - F_{\nu \mu \, })$ {\it and} that the first group of the Maxwell 
equations is satisfied:
\footnote{
Here, we extend the standard version of the Maxwell equations to this scalar theory of gravitation. Thus we do not discuss, for example, the fractional electromagnetic equations \cite{Baleanu2009}. The non-local character associated with fractional operators \cite{Baleanu2010} should be relevant to materials with e.m. memory properties \cite{Baleanu2009}.
}
\be\label{Maxwell 1}
F_{\lambda \mu \, ,\nu } + F_{\mu \nu ,\lambda 
} + F_{\nu \lambda ,\mu } = F_{\lambda \mu \, ;\nu } + F_{\mu \nu ;\lambda 
} + F_{\nu \lambda ;\mu } = 0.
\ee 
The physical significance of the field $\Mat{F}$ appears clearly with 
the Lorentz force ${\bf F}$ on a test particle with charge $q$. We have to 
find the expression of the Lorentz force in the presence of gravitation, 
subject to the conditions that (i) it is a space vector, invariant by the 
transformations (\ref{spatial change +f(t)}), and (ii) when the gravitational field vanishes, that expression must reduce to the following one, obtained in SR from the expression \cite{L&L} of the 4-force:
\be\label{Lorentz force SR}
F^i\equiv \frac{\dd}{\dd t}\left (m_0 \gamma _v \,\frac{\dd x^i}{\dd t} \right ) = \frac{\dd \tau }{\dd t} \frac{\dd \left ( m_0 c\,U^i\right )}{\dd \tau } = \frac{1}{\gamma _v}q F^i_{\ \, \mu }\,U^\mu ,
\ee
thus 
\be\label{Lorentz force SR-2}
F^i = q \left ( F^i_{\ \, 0 } +  F^i_{\ \, j }\,\frac{v^j}{c} \right).
\ee
In the present theory, the gravitational field determines the metric $\Mat{\gamma} $ and conversely (see Ref. \cite{A35}): there is no gravitation in some domain of spacetime, if and only if the metric is Minkowskian in that domain. Now the expression of the Lorentz force must depend only on the metric components at the given point 
of spacetime, not of their derivatives, because these derivatives make 
inertial or gravitational forces (and because it seems obvious that we need 
only the metric components in order to generalize (\ref{Lorentz force SR-2})). Hence, condition (ii) says that the Lorentz force must be written as Eq. (\ref{Lorentz force SR-2}) when the metric components are Minkowskian \textit{at the point considered}, ($\gamma_{\mu \nu })$ $= (\eta_{\mu \nu })$ $\equiv $ diag (1, $-$1, $-$1, $-$1). But, at the given point, due to the condition $\gamma_{0 i}=0$, the metric may always be set to the Minkowskian form \textit{by a coordinate change} \textit{(\ref{spatial change +f(t)})}. 
\footnote{\
By assumption: $\Mat{\gamma }$ is Lorentzian; $\gamma_{0 i}=0$; and $\gamma _{00}=\beta ^2>0$. Thus, the matrix $G\equiv (-\gamma _{ij})$ has Euclidean signature. Hence it can be put in the form $\mathrm{diag}(1,1,1)$ at any given ${\bf x}\equiv (x^i)$, by a  change (\ref{spatial change +f(t)})$_2$. We get  $\gamma' _{00}=(\frac{\dd x^0}{\dd x'^0})^2 \gamma _{00}=1$ at $x^0$, by a change (\ref{spatial change +f(t)})$_1$.
}
Now it is obvious that the following expression:
\be\label{Lorentz force ETG}
F^i = q \left ( \frac{F^i_{\ \, 0 }}{\beta } +  F^i_{\ \, j }\,\frac{v^j}{c} \right) = \frac{q}{c}\,F^i_{\ \, \mu }\,\frac{\dd x^\mu}{\dd t_{\bf x}} = \frac{q}{\gamma _v}\,F^i_{\ \, \mu }\,U^\mu  ,
\ee
gives the components of a space vector which is invariant under the transformations
(\ref{spatial change +f(t)}), and which reduces to (\ref{Lorentz force SR-2}) when ($\gamma_{\mu \nu })= (\eta _{\mu \nu })$ (i.e., when $\beta =1$ and $v^j =\dd x^j/\dd t$). 
Hence, Eq. (\ref{Lorentz force ETG}) is the general expression of the Lorentz force in the 
preferred frame assumed by the theory. It may be rewritten in space-vector 
form as
\be\label{Lorentz force ETG-vector}
{\bf F} = q \left ( {\bf E} +  {\bf v}\wedge \,\frac{{\bf B}}{c} \right),\quad \quad ({\bf a}\wedge {\bf b})^i \equiv e  ^i_{\ \, jk}\,a^j\,b^k,
\ee
where the electric and magnetic vector fields are the spatial vector fields with components 
\be\label{E and B}
E^i\equiv \frac{F^i_{\ \,0}}{\beta }, \quad B^k\equiv -\frac{1}{2}e^{ijk} F_{ij}.
\ee
In Eqs. (\ref{Lorentz force ETG-vector}) and (\ref{E and B}), $e_{ ijk}$ is the usual antisymmetric spatial tensor, its indices being raised or lowered using the spatial metric $\Mat{g}$ in the frame $\mathcal{E}$; in spatial coordinate systems whose natural basis is direct, we have $e_{ ijk}=\sqrt{g}\, \varepsilon_{ ijk} $, with $\varepsilon _{ ijk}$ the signature of the permutation $(i\,j\,k)$; it follows from the Leibniz formula for a determinant that we have then $e^{ ijk}=\frac{1}{\sqrt{g}}\, \varepsilon_{ ijk} $ \cite{L&L}.  The same expression (\ref{Lorentz force ETG}) is found if one asks that the charged particle must have 
the same 4-acceleration as in a metric theory, when the metric is\textit{ constant} in the reference fluid $\mathcal{E}$ (thus $g_{ij, 0} =0$). (For a metric theory, one merely applies the rule ``$\eta 
_{\mu \nu }$ becomes $\gamma_{\mu \nu }$ and comma goes to semi-colon" 
\cite{Will1993}.) This assumption is consistent with the fact that, for such a 
constant gravitational field, the dynamical equation (\ref{Eq T-f}) is the same as in 
metric theories, at least if there is no external force. In other words, for 
a gravitational field that is constant in the preferred frame, the present 
theory admits Einstein's equivalence principle in the classical form: ``in 
a local freely falling frame, the laws of non-gravitational physics are the 
same as in SR".\\

Considering now a continuous charged medium, we define $\rho 
_\mathrm{el} \equiv \delta q/\delta V$ and $J^{\, \mu } \equiv \rho_\mathrm{el}\,\dd x^\mu /\dd t_{\mathbf{x}}$. The Lorentz force density is written, in accordance with (\ref{Lorentz force ETG}), as
\be\label{Lorentz force density}
f^i \equiv \frac{\delta F^i}{\delta V} =F^i_{\ \, \mu }\,\frac{J^\mu}{c}.
\ee
The dynamical equation for the charged continuum is Eq. (\ref{Eq T-f}), with $f^i$ from Eq. (\ref{Lorentz force density}), and with the energy-momentum tensor ${\bf T}_\mathrm{charged\ medium}$ in the place of ${\bf T}$. Using the relations $F_{\mu \nu } = - F_{\nu \mu}$ and $\gamma_{0 i} =0$, one deduces from Eq. (\ref{Lorentz force density}), after a short algebra, the following expression for a term in Eq. (\ref{Eq T-f}):
\be\label{Lorentz force power}
\frac{{\bf f.v}}{c\,\beta } =F^0_{\ \, \mu }\,\frac{J^\mu}{c}.
\ee
In view of (\ref{Lorentz force density}) and (\ref{Lorentz force power}), we may write Eq. (\ref{Eq T-f}) for the charged medium as:
\be\label{Eq T-charged}
T^{\mu  \nu}_{\mathrm{charged\ medium}\ \ ;\nu} =b^\mu ({\bf T_\mathrm{charged\ medium}})+F^\mu_{\ \, \nu }\,\frac{J^\nu}{c}.
\ee
On the other hand, the \textit{total} energy-momentum is the sum ${\bf T} = 
{\bf T}_\mathrm{charged\ medium} + {\bf T}_\mathrm{field}$, with ${\bf T}_\mathrm{field}$ the 
energy-momentum tensor of the electromagnetic field \cite{L&L, Fock1964}:
\be\label{T em}
T_\mathrm{field}^{\, \mu \nu } \equiv \left (- F^\mu_{\ \ \lambda } F^{\, 
\nu \lambda } + \frac{1}{4}\gamma^{\, \mu \nu } F_{\lambda 
\rho } F^{\lambda \rho \, } \right)/4\pi . 
\ee
The total tensor {\bf T} obeys the general equation (\ref{Eq T-f}) for continuum 
dynamics (without any non-gravitational force, of course):
\be\label{Eq T}
T^{\mu  \nu}_{\ \ ;\nu} =b^\mu ({\bf T}).
\ee
Both the l.h.s. and the r.h.s. of this equation are linear in ${\bf T}$; see Eq. (\ref{b^mu}) for the r.h.s. Hence we may combine it with Eq. (\ref{Eq T-charged}) to {\it deduce} immediately that
\be\label{Eq T-field}
T^{\mu  \nu}_{\mathrm{field}\ \ ;\nu} =b^\mu ({\bf T_\mathrm{field}})-F^\mu_{\ \, \nu }\,\frac{J^\nu}{c}.
\ee
In words: \textit{the electromagnetic field may be considered as a ``material" continuum subjected to the gravitation and to the opposite of the Lorentz force}. We now show that\textit{ this gives the gravitational modification of Maxwell's second group in the present theory}. We first note that, using the antisymmetry of $\Mat{F}$ and the fact that $\gamma _{\mu \nu ;\rho }=0$, the definition (\ref{T em}) leads to 
\be\label{Eq T-field-2}
4 \pi T_{\mathrm{field}\ \,;\nu}^{\mu  \nu} =-F^\mu_{\ \ \lambda }\,F^{\nu \lambda }_{\ \ \,;\nu }+\frac{1}{2}F_{\nu \lambda } \left (F^{\nu \lambda ;\mu }+F^{\lambda \mu ;\nu }+F^{\mu \nu ;\lambda }\right )= -F^\mu_{\ \ \lambda }\,F^{\nu \lambda }_{\ \ ;\nu }.
\ee
(The second equality is a consequence of the first group, Eq. (\ref{Maxwell 1}), using again $\gamma _{\mu \nu ;\rho }=0$ and some index shuffling.) Rewriting Eq. (\ref{Eq T-field}) with the help of Eq. (\ref{Eq T-field-2}), we get
\be\label{Eq T-field-3}
F^\mu_{\ \ \lambda }\,F^{\lambda \nu }_{\ \ \,;\nu }= 4 \pi b^\mu \left ({\bf T}_\mathrm{field} \right)-4 \pi F^\mu_{\ \ \lambda }\,\frac{J^\lambda }{c},
\ee
where $b^\mu \left ({\bf T}_\mathrm{field} \right)$ is given by Eqs. (\ref{b^mu}) and (\ref{T em}). This gives the second group of the Maxwell equations in the 
presence of gravitation, according to the investigated theory --- at least 
for the generic case where the field tensor $\Mat{F}$ is invertible (det 
$\Mat{F}\equiv $ det ($F^\mu _{\ \ \nu}) \ne 0$). Indeed, 
multiplying on the left by the complementary matrix $(\widetilde{F}^\rho _{\ \ \mu })$, we 
obtain:
\be\label{Eq T-field-4}
(\mathrm{det}\, \Mat{F}) F^{\rho   \nu } _{\ \ \,;\nu  }= 4 \pi \left (\widetilde{F}^\rho   _{\ \ \mu  }\,b^\mu  ({\bf T}_\mathrm{field}) - (\mathrm{det}\, \Mat{F}) \frac{J^\rho   }{c}\right),
\ee
which is rewritten, for an invertible matrix ($F^\mu _{\ \ \nu })$, as
\be\label{Maxwell ETG}
F^{\mu \nu } _{\ \ \,;\nu  }= 4 \pi \left (G^\mu  _{\ \ \nu }\,b^\nu ({\bf T}_\mathrm{field}) - \frac{J^\mu  }{c}\right),\quad (G^\mu  _{\ \ \nu })\equiv (F^\mu  _{\ \ \nu })^{-1}.
\ee
(Note that $\Mat{G}$, like $\Mat{F}$, is an antisymmetric tensor, $G_{\nu \mu } = - G_{\mu \nu }$.)

\section{Comments on the gravitationally-modified Maxwell equations in the 
theory}

{\bf i}) If the gravitational field is constant in the preferred reference fluid, 
whence $g_{ij, 0} =0$, it results from (\ref{b^mu}) that Eq. (\ref{Maxwell ETG}) reduces to the gravitationally-modified second group in GR and other metric theories of gravitation:
\be\label{Maxwell GR}
F^{\mu \nu } _{\ \ \,;\nu  }= -4 \pi \frac{J^\mu  }{c}.
\ee
In general, however, $g_{ij, 0} \ne 0$, in which case the modified 
second group remains in the \textit{non-linear} form (\ref{Maxwell ETG}). But the Maxwell equations (\ref{Maxwell GR}) are {\it also} non-linear in metric theories, e.g. in GR, because the energy-momentum of the electromagnetic field does contribute to the r.h.s. of the non-linear Einstein equations. That is, the field 
$\Mat{F}$ influences the metric non-linearly
, whereas the modified Maxwell equations (\ref{Maxwell GR}) depend on the metric. This is also true in the investigated theory. In the latter, the second group of the modified Maxwell equations, Eq. (\ref{Maxwell ETG}), is more complex than Eq. (\ref{Maxwell GR}). Yet the metric is determined by just the scalar linear wave equation (\ref{wave eqn v2}), which is much simpler than the Einstein equations.\\

\noi {\bf ii}) Among local coordinate systems that are adapted to the reference 
fluid and in which the  synchronization (\ref{gamma_0i=0}) applies (two such systems exchanging by (\ref{spatial change +f(t)})), there are ones for which, at the event $X$ considered, the spatial natural basis $(\partial _i)$ is orthonormal for the space metric $\Mat{g}$, i.e., $g_{ij}(X)=\delta_{ij}$. We then get easily from (\ref{E and B}):
\be\label{F^mu_nu; g orthonormal}
F^i_{\ \,0}=\beta \,E^i,\quad F^0_{\ \, i}=\beta ^{-1}\,E^i,\quad F^i_{\ \,j}=\varepsilon _{ijk}\,B^k.
\ee
Furthermore, by rotating the axes we can get $B^{\, i} = B \delta_{1 i}$ and $E^{\, 3} =0$ at this event. In such coordinates, we get from (\ref{F^mu_nu; g orthonormal}): $\mathrm{det}(F^\mu_{\ \, \nu}) = -(E^1)^2 B^2$. Therefore, the condition $\mathrm{det}\,\Mat{F}$ $\ne $ 0 is equivalent to ${\bf E.B}\equiv \Mat{g}({\bf E},{\bf B})\ne 0$. (By the way, it can be shown using (\ref{E and B}) that we have generally ${\bf E.B} = - e^{\mu \nu \rho \psi } F_{\mu \nu } F_{\rho \psi } /8 $, where one defines $e^{\mu \nu \rho \psi }\equiv \frac{1}{\sqrt{-\gamma }} \varepsilon _{\mu \nu \rho \psi }$ as for $e^{i j k}$ after Eq. (\ref{E and B}) \cite{Stephani1982}.) The condition ${\bf E.B}\ne 0$ is satisfied at ``generic" points for a ``generic" electromagnetic field but, unfortunately, it is not satisfied by the simplest examples of such fields, viz. purely electric and purely magnetic fields, nor by ``simple" electromagnetic waves (since the characteristic property of the latter or ``null fields" is that both invariants are zero, ${\bf E.B} =0$ and ${\bf E}^{2} - {\bf B}^{2} = - F_{\mu \nu } F^{\, \mu \nu }/2 = 0$). One may show, however, that for a field which is purely electric (in the preferred reference fluid), the new term in the r.h.s. of Eq. (\ref{Maxwell ETG}), $a^{\, \mu }\equiv 4 \pi \widetilde{F}^\mu  _{\ \ \nu }\,b^\nu ({\bf T}_\mathrm{field}) /(\mathrm{det}\, \Mat{F})$, can be defined by a continuity extension. Yet this is not the case for a purely magnetic field or for a simple wave. On the other hand, the necessary restriction to invertible field tensors could present a problem only if, in some physically reasonable situation with a variable gravitational field, Eq. (\ref{Maxwell ETG}) would have no solution (i.e., no invertible solution). It is likely that the coupling with a variable gravitational field would forbid to have such peculiar fields as, for instance, null fields, purely electric or magnetic fields as exact solutions of Eq. (\ref{Maxwell ETG}) --- although, for the weak and slowly varying gravitational fields which we live in, solutions very close to such fields could exist. Moreover, we may consider the simple waves as physically important model fields, which are clearly not solutions of Eq. (\ref{Maxwell ETG}) for a variable gravitational field, but to which Eq. (\ref{Eq T-field-3}) still applies (as well as Eq. (\ref{Eq T-field-4})). As it will be seen in Section \ref{Wave vs ray} below, Eq. (\ref{Eq T-field-3}) is enough for the transition to geometrical optics.\\

\noi {\bf iii}) This theory is a classical one, hence a macroscopic one. Remind that the charged medium is modelled as a continuous extended object, not as a set of point particles. The Maxwell equations (\ref{Eq T-field-3}) have to be considered exact in the framework of this theory, hence no radiation reaction has to be added (except, possibly, as a step in an iterative approximation process to solve the equations). The combined motion of the charged continuous medium and the e.m. field (including e.m. radiation) should be obtained, in principle, by solving together: the dynamical equation (\ref{Eq T-charged}), the state equation for the charged continuum, and Eqs. (\ref{Maxwell 1}) and (\ref{Eq T-field-3}). Note that there is an exact local energy conservation equation for the most general case in this theory: Eq. (30) in Ref. \cite{A35}. This equation does apply here, in view of Eq. (\ref{Eq T}). \\ 

\noi {\bf iv}) It is well known that Eq. (\ref{Maxwell GR}) implies the charge 
conservation, which passes thus from SR to GR. In the present theory, we use 
the antisymmetry of $\Mat{F}$ as in GR, so that $F^{\mu \nu } _{\ \,;\nu;\mu   }=0$, and 
get from Eq. (\ref{Maxwell ETG}):
\be\label{Charge rate ETG}
\hat{\rho  } \equiv \left ( J^\mu \right )_{; \mu}= c \left( G^\mu  _{\ \,\nu }\,b^\nu ({\bf T}_\mathrm{field}) \right )_{; \mu},
\ee
according to which the charge of the continuum is not exactly conserved if the gravitational field varies in the ether frame (recall from (\ref{b^mu}) that $b^\nu=0$ if it does not). In  previous works \cite{A35,A20}, we found that matter may be produced or destroyed, its rest-mass energy being taken from or given to the variable gravitational field. In the same way, we find now that, depending on the orientation of the electromagnetic field relative to the variation of the gravitational field, the charge of a continuous distribution may ``locally" vary --- though rather at a macroscopic scale, since here we are discussing a classical continuous medium, not (quantum) elementary particles. (Of course, this would imply that elementary charges are produced or destroyed in a variable gravitational field, but a classical theory cannot describe how; the same is true for matter production/destruction.) In an electromagnetic wave, the inverse field $\Mat{G}$, as well as the field  $\Mat{F}$, alternate rapidly, whereas the $b^{\nu }$ 's keep the same sign, so that the charge balance is nearly zero over one period. The same cannot be said for a slowly varying field such as the Earth's magnetic field, but one has also to take into account the variation of the gravitational field. For the mass balance, it was easy to give reasonable estimates, which turn out to be very tenuous and hence seem compatible with the experimental evidence on ``mass conservation" \cite{A35,A20}. In contrast, we do not find straightforward to assess the amount of the charges that should be produced in realistic situations, in application of Eq. (\ref{Charge rate ETG}). This is partly due to the relative complexity of the terms in Eq. (\ref{Charge rate ETG}), partly also to the difficulty mentioned at point {\bf ii}) hereabove --- that the simplest solutions are not usable. Hence, we leave the obtainment of such estimates to a future numerical work. Clearly, we expect that the amounts are very small in usual situations. 

\section{Dynamical link between wave optics and ray optics under the 
gravitation}\label{Wave vs ray}

In the present theory, as also in GR and in other relativistic theories of 
gravitation, the electromagnetic rays are defined as the trajectories of 
light-like particles (i.e., $\dd s^2 =0$) called photons --- a photon being 
defined by its energy $E$ or its frequency $\nu $, related by $E = h\nu $. In 
GR, these trajectories are geodesic lines, whereas here they are governed 
(\textit{in vacuo}) by ``Newton's second law" (\ref{Newton 2nd law}) (with ${\bf F}={\bf 0} $). As this photon dynamics is for zero external force except gravitation, the search of a direct, dynamical link between electromagnetic field and photons trajectories leads us to examine in which conditions the electromagnetic field may be seen as a ``dust of photons". Thus, the energy-momentum tensor (\ref{T em}) should have the form of the tensor for (ordinary) dust. Is that possible? At least, we may ask that the energy-momentum tensor of the electromagnetic field have the form
\be\label{T = tensor product}
T^{\mu \nu } = V^\mu  V^\nu , 
\ee
which is generally covariant. For instance by selecting coordinates adapted to the preferred reference fluid and such that, at the event considered: the spatial natural basis $(\partial _i)$ is orthonormal for $\Mat{ g}$; $B^i= B \delta_{1 i}\,$; and $\,E^3=0$ --- one verifies easily that the {\it necessary and sufficient condition} for {\bf T}$_\mathrm{field}$ to have the 
form (\ref{T = tensor product}) is that both invariants of the field tensor $\Mat{F}$ must be 
zero (see also Stephani \cite{Stephani1982}), thus a ``null field". \\

On the other hand, in the preferred reference fluid, we may rewrite (\ref{T = tensor product}) in the same form that we used for ordinary dust, Eq. (\ref{T dust-2}) --- which does not determine uniquely $u^{\, \mu }$ and $\rho $. Yet if we want to use the relation $u^{\, 0} \equiv c$, as we did for ordinary dust, then we must fix $\rho \equiv T^{\, 00}/(\tilde{\beta }^2 c^{2})$ (as we did for ordinary dust, Eq. (\ref{Def rho v2})). This determines also the spatial components: $u^{\, i} \equiv c T^{i 0}/T^{00}$, Eq. (\ref{Current free dust}). Then, starting from (\ref{Newton 2nd law continuum}), all equations of Subsect.  \ref{Newton continuum} apply (except for (\ref{T dust-1})). Thus, \textit{independently of the exact physical nature of the material or field}, the fact that the energy-momentum tensor has the form (\ref{T = tensor product}) ensures that the dynamical equation (\ref{Eq T-f}), with the external force field {\bf f} $=$ 0, is equivalent to the conjunction of Eqs. (\ref{Newt 2nd continuum-4}), with $f '_{\, i} =$ 0, and (\ref{Energy free dust-2}). (These equations are valid in coordinates such that $(g^{\, 0})_{, j} = 0$, and with $x^{\, 0} =$ \textit{cT} where $T$ is the ``absolute time".) The absolute 3-velocity field ${\bf u} = \dd {\bf x}/\dd T $ of the continuous medium is defined by Eq. (\ref{Current free dust}): it is the velocity of the energy flux, relative to the preferred reference fluid. In turn, Eqs. (\ref{Newt 2nd continuum-4}) and (\ref{Energy free dust-2}) are the respective exact translations, in these adapted coordinates, of Newton's second law for the continuous medium, Eq. (\ref{Newton 2nd law continuum}) with the non-gravitational force ${\bf f '} = 0$, and the 
energy equation (\ref{Energy free dust-0}).\\

Let us come back to the special case of the electromagnetic field \textit{in vacuo}. In this 
case, the dynamical equation (\ref{Eq T-f}) is nothing else than the modified second 
group (\ref{Eq T-field-3}) of Maxwell equations, with $J^\lambda  =0$:
 \be\label{Eq T-field-5}
F^\mu_{\ \ \lambda }\,F^{\lambda \nu }_{\ \ ;\nu }= 4 \pi b^\mu \left ({\bf T}_\mathrm{field} \right).
\ee
Hence, the foregoing means that, for a null field $\Mat{F}$ and \textit{only} for a null 
field, these electromagnetic equations\textit{ in vacuo} are exactly equivalent to Newton's 
second law (\ref{Newton 2nd law continuum}) for the electromagnetic dust (i.e., ${\bf f'}=0$) and the corresponding energy equation, Eq. (\ref{Energy free dust-0}). And 
this is indeed a dust made of light-like particles or photons, because the 
absolute velocity {\bf u} of the dust is defined by Eq. (\ref{Current free dust}), so that 
the velocity measured with physical clocks (bound to the preferred reference fluid, 
but affected by the gravitational field), ${\bf v} \equiv  
\dd {\bf x}/\dd t_{\mathbf{x}} = {\bf u}/\tilde{\beta}$, satisfies
\bea
c^2-g_{ij}\,v^i\,v^j & = & c^2 \left (1 - \frac{g_{ij}\,T^{i 0}\,T^{j 0}}{\tilde{\beta } ^2 (T^{0 0})^2} \right )= c^2 \left (1 - \frac{g_{ij}\,V^i\,V^j}{\tilde{\beta } ^2 \,V^0\,V^0} \right )\nonumber \\
& = & c^2 \left (\frac{V_0\,V^0+V_i\,V^i}{V_0\,V^0} \right ) = c^2 \frac{T^\mu_{\ \,\mu}}{T^0_{\ \,0}},
\eea
which is nil for the energy-momentum tensor of any electromagnetic field. 
\textit{In summary:} when {\bf T}$_\mathrm{field}$ is the energy-momentum tensor associated with a \textit{null} electromagnetic field $\Mat{F}$, and only then, Eq. (\ref{Eq T-field-5}) says exactly: 

(i) that the trajectories {\bf x}($t)$ which are defined by ${\bf u} = \dd {\bf x}/\dd t $ with ${\bf u}$ deduced from ${\bf T}_\mathrm{field}$ by $T^{0 0}\,u^i=cT^{i 0}$, are photon trajectories --- that is, trajectories defined by Newton's second law (\ref{Newton 2nd law}) applied to a free light-like particle;

(ii) that one has the continuous form for dust, Eq. (\ref{Energy free dust-0}), of the energy equation (\ref{eq3}). (For a test particle, the energy equation is a consequence of Newton's second law, but this does not hold true for a continuum.)\\

A comment may be in order, to make clearer what are photon trajectories in the present theory. Equation (\ref{Newton 2nd law}) involves the energy $E$, which varies along the trajectory according to Eq. (\ref{eq3}). The only difference between mass 
points and photons which is relevant to this dynamics is that photons are 
light-like particles. Using Eq. (\ref{Newton 2nd law}) with ${\bf F} ={\bf 0}$, together 
with the definition of the velocity ${\bf v}$ and its modulus $v$ (Eq. (\ref{Def v}))), 
the assumption $v = c$ leads to the differential system
\be\label{Dyn v=c}
\frac{D {\bf v}}{Dt} = \beta \left ( {\bf g}-({\bf g.v})\,\frac{{\bf v}}{c^2} \right ),\quad {\bf v}=\frac{1}{\beta } \frac{ \dd {\bf x}}{\dd t}.
\ee
It may be proved that any solution of (\ref{Dyn v=c}) with an initial data such that 
$v(t_{\, 0}) = c$, satisfies $v = c$ at any time. Thus, photons trajectories 
are just solutions of the system (\ref{Dyn v=c}) with any initial data such that 
$v(t_{\, 0}) = c$ (evaluated with local standards according to Eq. (\ref{Def v})). Of 
course, this same equation (\ref{Dyn v=c}) is deduced if one starts from the 
continuous form for dust, Eq. (\ref{Newton 2nd law continuum}) with ${\bf f'} = 0$, 
because one just has to substitute $\delta E$ for $E$.

\section{Discussion and conclusions}

{\bf i}) The present theory derives the motion of test particles from a unique extension of the relativistic form of Newton's second law to any given reference fluid in a spacetime curved by gravitation \cite{A16}. More precisely, the theory assumes that there is a preferred reference fluid $\mathcal{E}$, in which, just as in Newtonian theory, the gravity acceleration {\bf g} derives from a spatial potential: $U \equiv -c^{\, 2}\Log \beta $ with $\beta \equiv \sqrt{(\gamma _{0 0})_\mathcal{E}}$, where $(\gamma _{0 0})_\mathcal{E}$ is the $\gamma _{0 0}$ component of the spacetime metric $\Mat{\gamma }$ (in coordinates adapted to the reference fluid $\mathcal{E}$ and such that the synchronization condition (\ref{gamma_0i=0}) applies). Here, equations governing the dynamics of a general continuum in the presence of gravitational and non-gravitational forces have been derived by two 
independent methods, by induction from the case of a test particle. The two 
methods give the same result for a dust. For a general energy-momentum 
tensor, the comparison of the two methods gives access to the internal 
forces in the continuous medium.\\

\noi {\bf ii}) The electromagnetic field tensor $\Mat{F}$ is stated to derive 
from a 4-potential in the usual way. The expression of the Lorentz force 
is derived uniquely from the requirement that it is a space vector which 
must reduce to the classical expression in the absence of gravitation. The 
total energy-momentum tensor {\bf T} is assumed to be the sum of the 
classical tensor {\bf T}$_\mathrm{charged\ medium}$ for the charged continuum and 
the classical tensor {\bf T}$_\mathrm{field}$ for the electromagnetic field. It 
must obey the general equation for continuum dynamics in the absence of 
non-gravitational external force. On the other hand, the tensor 
{\bf T}$_\mathrm{charged\ medium}$ must obey the general equation for continuum dynamics in 
the presence of the Lorentz force due to the electromagnetic field. This 
determines uniquely the form taken by the Maxwell equations in a 
gravitational field, according to the present theory: Eq. (\ref{Eq T-field-3}) or (equivalently in general) Eq. (\ref{Maxwell ETG}).\\

\noi {\bf iii}) In the special case that the field tensor $\Mat{F}$ makes a singular $4 \times 4$ matrix, and if moreover the gravitational field is variable, then only the form (\ref{Eq T-field-3}) of the modified Maxwell equations is valid. The form (\ref{Eq T-field-3}) is yet sufficient to make the transition from wave optics to geometrical optics \textit{in vacuo}, in the presence of gravitation. The transition consists essentially in showing: {\it a}) that a continuous distribution of ``free" photons can be defined as an ``electromagnetic dust", i.e., as a continuum whose energy-momentum tensor is given by the usual expression for an electromagnetic field (\ref{T em}), {\it and} which obeys Newton's second law for a continuous medium subjected only to the gravitational force [Eq. (\ref{Newton 2nd law continuum}) with ${\bf f '} = 0$], plus the energy equation (\ref{Energy free dust-0}). {\it b}) That, in this case, these two dynamical laws are nothing else than the modified Maxwell equations {\it in vacuo,} Eq. (\ref{Eq T-field-5}). {\it c}) That, for such an electromagnetic dust, the modified Maxwell equations imply that each trajectory of the energy flux is indeed a photon trajectory of the present theory.\\

\noi {\bf iv}) The modified Maxwell equations also imply that the
(macroscopic) conservation of the electric charge of a continuous medium would be violated in a variable gravitational field, according to that theory in its present state. In a past work, 
it had already been found that mass conservation would be violated in a 
variable gravitational field, and it had been shown that the amounts are 
extremely tenuous in usual conditions and so remain compatible 
with the experimental evidence on mass conservation \cite{A35}. Unfortunately, 
the calculations for the charge balance are more involved, so that no 
estimate seems to be easily obtainable without having recourse to a 
numerical work. This is obviously a crucial and dangerous point for the 
theory, but the present work has been focused on the internal consistency 
of the theory: the important point, in this respect, is that the Maxwell 
equations are derived unambiguously, and are consistent with photon 
dynamics. We noted that the theory predicts a macroscopic charge production/destruction but does not say which kind of particles and which elementary processes could be involved. Therefore, the strong experimental evidence for the absence of the so-far investigated charge-conservation-violating decays, e.g. electron decays \cite{Agostini-et-al.2015}, does not prove that such a production is excluded. Moreover, as for mass production/destruction, one may argue that charge production/destruction should be allowed in a cosmological context, and that a cosmological context should be only a particular case for a theory \cite{A20}. Thus, the charge non-conservation in the present theory will become an interesting feature, if it turns out to be negligible in situations where charge is indeed experimentally found to be conserved. In particular, since the gravitational field of an astronomical object varies in its translation through the imagined ``ether", charge production/destruction in a varying gravitational field should have some implications on the magnetic fields of the astronomical objects, which are not fully understood. We hope to be able to investigate this in the future.\\

\noi {\bf Acknowledgement.} I am grateful to Gonzalo Ares de Parga for a useful discussion.



\begin{thebibliography}{9}
\small

\bibitem{Poincare1904}
Poincar\'{e} H., L'\'etat actuel et l'avenir de la physique math\'ematique, Bull. Sci. Math. (S\'{e}r. 2), 1904, 28, 302-324.

\bibitem{Poincare1905}
Poincar\'{e} H., Sur la dynamique de l'\'electron, C.--R. Acad. Sci., 1905, 140, 1504-1508.

\bibitem{Poincare1906}
Poincar\'{e} H., La dynamique de l'\'electron, 
Rendic. Circ. Matemat. Palermo, 1906, 21, 129-176.

\bibitem{Builder1958a}
Builder G., Ether and relativity, Austr. J. Phys., 1958, 11, 279-297.

\bibitem{Builder1958b}
Builder G., The constancy of the velocity of light, Austr. J. Phys., 1958, 11, 457-480.

\bibitem{Janossy1965}
J\'{a}nossy L., The Lorentz principle, Acta Phys. Polon., 1965, 27, 61-87.

\bibitem{Prokhovnik1993}
Prokhovnik S.J., The physical interpretation of special relativity -a vindication of Hendrik Lorentz, Z. Naturforsch., 1993, 48a, 925-931.

\bibitem{Prokhovnik1967}
Prokhovnik S.J., The logic of special relativity, Cambridge University Press, Cambridge, U.K., 1967

\bibitem{O3}
Arminjon M., Gravity as Archimedes' thrust and a bifurcation in that theory, Found. Phys., 2004, 34, 1703-1724.

\bibitem{A18}
Arminjon M., Scalar theory of gravity as a pressure force, Rev. Roum. Sci. Tech.- M\'{e}c. Appl., 1997, 42, 27-57. 

\bibitem{A35}
Arminjon M., Space isotropy and weak equivalence principle in a scalar theory of gravity, Braz. J. Phys., 2006, 36, 177-189.

\bibitem{Will1993}
Will C.M., Theory and experiment in gravitational physics, 2nd ed., Cambridge University Press, Cambridge, U.K., 1993

\bibitem{JacobsonMattingly2001}
Jacobson T., Mattingly D., Gravity with a dynamical preferred frame, Phys. Rev. D, 2001, 64, 024028.

\bibitem{A28}
Arminjon M., Accelerated expansion as predicted by an ether theory of gravitation, Phys. Essays, 2001, 14, 10-32.

\bibitem{FriedrichRendall2000}
Friedrich H., Rendall A. D., The Cauchy problem for the Einstein equations, In: Schmidt B.G. (Ed.), Einstein's field equations and their physical implications, Lecture Notes in Physics No. 540, Springer-Verlag, Berlin - Heidelberg, 2000, 127-224

\bibitem{A48}
Arminjon M., A simpler solution of the non-uniqueness problem of the covariant Dirac theory, Int. J. Geom. Meth. Mod. Phys., 2013, 10, No. 7, 1350027.  

\bibitem{B26}
Arminjon M., Scalar gravity with preferred frame: asymptotic post-Newtonian scheme and the weak equivalence principle, In: Fiziev P., Todorov M. (Eds.), Gravity, astrophysics and strings at the Black Sea: second advanced research workshop, St. Kliment Ohridski University Press, Sofia, Bulgaria, 2005, 1-16

\bibitem{A16}
Arminjon M., On the extension of Newton's second law to theories of gravitation in curved space-time, Arch. Mech., 1996, 48, 551-576.

\bibitem{WattMisner1999}
Watt K., Misner C.W., Relativistic scalar gravity: a laboratory for numerical relativity, Preprint arXiv:gr-qc/9910032, 1999.

\bibitem{Ni1972} 
Ni W.-T., Theoretical frameworks for testing relativistic gravity. IV. A compendium of metric theories of gravity and their post-Newtonian limits, Astrophys. J., 1972, 176, 769-796.

\bibitem{O2}
Arminjon M., Ether theory of gravitation: why and how? In: Duffy M.C., Levy J. (Eds.), Ether, space-time and cosmology, Vol. 1: Modern ether concepts, relativity and geometry,  PD Publications, Liverpool, U.K., 2008, 139-201

\bibitem{A19}
Arminjon M., Post-Newtonian approximation of a scalar theory of gravitation and application to light rays, Rev. Roum. Sci. Tech.- M\'{e}c. Appl., 1998, 43, 135-153.

\bibitem{A23}
Arminjon M., Asymptotic expansions for relativistic celestial mechanics, Roman. J. Phys., 2000, 45, 389-414.

\bibitem{B21}
Arminjon M., The scalar ether-theory of gravitation and its first test in celestial mechanics, In: Mostepanenko V.M., Romero C. (Eds.), Proc. 5th Friedmann international seminar on gravitation and cosmology: Int. J. Mod. Phys., 2002, A17, 4203-4208. 

\bibitem{A34}
Arminjon M., Gravitational effects on light rays and binary pulsar energy loss in a scalar theory of gravity, Theor. Math. Phys., 2004, 140, 1011-1027 [Teor. Mat. Fiz., 2004, 140, 139-159]. 

\bibitem{Everitt2011}
Everitt C.W.F., DeBra D.B., Parkinson B.W., Turneaure J.P., et al., Gravity Probe B: final results of a space experiment to test general relativty, Phys. Rev. Lett., 2011, 106, 221101.

\bibitem{Schiff1960}
Schiff L.I., Possible new experimental test of general relativity theory, Phys. Rev. Lett., 1960, 4, 215-217.

\bibitem{Breakwell1988}
Breakwell J.V., The Stanford relativity gyroscope experiment: correction to the predicted geodetic precession of the gyroscope resulting from the Earth's oblateness, In: Fairband J.D., et al. (Eds.), Near zero: new frontiers of physics, Freeman, New York, 1988, 685-690

\bibitem{A36}
Arminjon M., Equations of motion according to the asymptotic post-Newtonian scheme for general relativity in the harmonic gauge, Phys. Rev. D, 2005, 72, 084002.


\bibitem{Stephani1982}
Stephani H., General relativity --- an introduction to the theory of the gravitational field, Cambridge University Press, Cambridge, U.K., 1982

\bibitem{A15}
Arminjon M., Energy and equations of motion in a tentative theory of gravity with a privileged reference frame, Arch. Mech., 1996, 48, 25-52.

\bibitem{A20}
Arminjon M., On the possibility of matter creation/destruction in a variable gravitational field, Analele Universit. Bucure\c{s}ti -- Fizic\u{a}, 1998, 47, 3-21.

\bibitem{Lichne1962}
Lichnerowicz A., In: Lichnerowicz A., Tonnelat M.-A. (Eds.), Les th\'{e}ories relativistes de la gravitation, Ed. du Centre National de la Recherche Scientifique, Paris, 1962, 93-106

\bibitem{Synge1964}
Synge J.L., In: De Witt B., De Witt C. (Eds.), Relativity, groups and topology, Gordon and Breach, New York - London, 1964, 79-88

\bibitem{deFelice-Clarke1990}
de Felice F., Clarke C.J.S., Relativity on curved manifolds, Cambridge University Press, Cambridge, U.K., 1990

\bibitem{Fock1964}
Fock V., The theory of space, time and gravitation, 2nd English edition, Pergamon, Oxford, U.K., 1964 

\bibitem{L&L}
Landau L.D., Lifshitz E.M., The classical theory of fields, 3rd English edition, Pergamon, Oxford, U.K., 1971 

\bibitem{A53}
Arminjon M., On the definition of energy for a continuum, its conservation laws, and the energy-momentum tensor, Adv. Math. Phys., 2016, 9679460.

\bibitem{Cattaneo1958}
Cattaneo C., General relativity: relative standard mass, momentum, energy and gravitational field in a general system of reference, Nuovo Cim., 1958, 10, 318-337.

\bibitem{RodriguesCapelas2007}
Rodrigues W.A. Jr., Capelas de Oliveira E., The many faces of Maxwell, Dirac and Einstein equations. A Clifford bundle approach, Lecture Notes in Physics No. 722, Springer International Publishing, Switzerland, 2007, p. 174

\bibitem{A52}
Arminjon M., Defining the space in a general spacetime, Int. J. Geom. Meth. Mod. Phys., 2016, 13, No. 3, 1650031.

\bibitem{Baleanu2009}
Baleanu D., Golmankhaneh Ali K., Golmankhaneh Alireza K., Baleanu M.C., Fractional electromagnetic equations using fractional forms, Int. J. Theor. Phys., 2009, 48, 3114-3123.

\bibitem{Baleanu2010}
Baleanu D., Golmankhaneh Alireza K., Nigmatullin R., Golmankhaneh Ali K., Fractional Newtonian mechanics, Open Physics, 2010, 8, No. 1, 120-125.

\bibitem{Agostini-et-al.2015}
Agostini M., et al. (Borexino Collaboration), A test of electric charge conservation with Borexino, Phys. Rev. Lett., 2015, 115, No. 23, 231802. 


\end{thebibliography}
\end{document}